\documentclass{article}
\usepackage{natbib}

\usepackage[a4paper, total={6in, 8in}]{geometry}
\usepackage{setspace}
\usepackage{tocbibind}
\usepackage{comment}
\usepackage{bm}
\usepackage{xcolor}
\usepackage{amsthm}
\usepackage{amssymb}

\newtheorem{theorem}{Theorem}

\newtheorem*{theorem*}{Theorem}
\newtheorem{lemma}[theorem]{Proposition}

\newtheorem{proposition}[theorem]{Proposition}

\theoremstyle{definition} 
\newtheorem{rem}{Remark}

 \theoremstyle{definition} 
 \newtheorem{exm}{Example}

\usepackage{hyperref}
\usepackage{graphicx}
\usepackage{helvet}

\usepackage{amsmath,amssymb,mathptmx,txfonts}

\usepackage{amsmath, amsfonts}

\begin{document} 

\title{\Large{\textbf{Nonlinearity in Dynamic Causal Effects: Making the Bad into the Good, and the Good into the Great? \thanks{We would like to thank Thomas Carr and an anonymous referee for beneficial comments. Kitagawa gratefully acknowledges the financial support from NSF through the NSF Standard Grant  (FAIN 234361). 
Weining Wang is supported through the project ``IDA Institute of Digital Assets'', CF166/15.11.2022,  financed under the Romania’s National Recovery and Resilience Plan; and the Marie Skłodowska-Curie Actions under the European Union's Horizon Europe research and innovation program for the Industrial Doctoral Network on Digital Finance, acronym: DIGITAL, Project No. 101119635.
} \\
}\Large }}
\author{Toru Kitagawa\thanks{Department of Economics, Brown University. Email: toru\_kitagawa@brown.edu} \hspace{0.5cm} Weining Wang\thanks{Department of Economics, University of Bristol. Email: weining.wang@bristol.ac.uk} \hspace{0.5cm} Mengshan Xu\thanks{Department of Economics, University of Mannheim. Email: mengshan.xu@uni-mannheim.de}}

\date{}
\maketitle
\begin{abstract}
    This paper was prepared as a comment on ``Dynamic Causal Effects in a Nonlinear
World: the Good, the Bad, and the Ugly" by Michal Koles\'{a}r and Mikkel Plagborg-M{\o}ller. 
We make three comments, including a novel contribution to the literature, showing how a reasonable economic interpretation can potentially be restored for average-effect estimators with negative weights.
\end{abstract}
\section{Background}%
\onehalfspacing
We are grateful for the opportunity to discuss this insightful paper by Michal Koles\'{a}r and Mikkel Plagborg-M{\o}ller (KP hereafter). For more than three decades, program evaluation, causal inference, and nonseparable structural equation models have been prominent areas of research in econometrics. These topics have interacted with the ``credibility revolution" in applied microeconomics, which has driven methodological advancements and strengthened the credibility of evidence obtained through empirical research. Key features of these approaches are (i) minimal restrictions on the heterogeneity of causal effects and the functional form of causal equations and (ii) providing a transparent interpretation for causal estimands in order to clarify their relevance and usefulness for policy analysis. KP shares these features, and we believe that it offers novel and useful perspectives on causal analysis in macroeconometrics.  

To motivate KP's approach, it is useful to contrast it with conventional econometric research, in which a causal model is combined with identification assumptions to yield a statistical estimand: 
\[
\mbox{(Causal model) + {(Identification)}} \Rightarrow \mbox{{(Statistical estimand)}}.
\]
A more recent contrasting approach, which we call ``reverse econometrics'', begins with a statistical estimand which is then combined with identification assumptions to obtain a causal interpretation:
\[
 \mbox{{(Statistical estimand)}+{(Identification)}} \Rightarrow \mbox{(Causal parameter)}.
\]
Research in reverse econometrics has been growing. Influential papers include \cite{imbens1994identification} which introduces the local average treatment effect interpretation of the two-stage least squares estimand, \cite{angrist1995estimating} which derives the weighted-average conditional average treatment effect interpretation of the ordinary least squares estimator with additive controls, and \cite{de2020two} which shows the weighted-average treatment effect interpretation of the two-way fixed effects estimator. A particular merit of the reverse econometrics approach is that it shows the robustness and policy relevance of existing, statistically well-studied estimands in the presence of heterogeneity or nonlinearity.
KP performs reverse econometric analysis of the impulse response (IR) estimands that are dominant in linear time-series models. See \cite{rambachan2021when}, \cite{casini2024identification}, and \cite{Koo_etal_2023} for other reverse econometric analyses of impulse response estimands; see also \cite{chen2024potential}, which seeks to offer alternative interpretations of linear causal estimands.

In the following, we make three points. First, we argue that the presence of negative weights is not always problematic. We present a novel proposition that shows conditions under which a ``bad'' estimand can be salvaged by transforming it from a weighted-average of structural derivatives with negative weights into one with positive weights without altering the estimand's value. Second, we question how nonstationary causal effects can be accommodated. Third, we discuss the possibility of elevating the ``good'' properties of the estimands to ``great'' by linking the reverse econometrics to policy decision making. Throughout our discussion, we maintain the definitions and notation of KP.

\section{Comment 1: Negative Weights} \label{sec:comment1}

 KP focuses on the estimation of the average marginal effect, which can be expressed as
\begin{equation} \label{eq:weigted_effect}
  \beta =\int_{\mathcal{X}} \omega(x) g^{\prime}(x) d x,
\end{equation}
where $g(x)$ represents the average of a structural function of randomly assigned shocks $x$, and $\omega(\cdot)$ is a weight function determined by the associated estimation procedure.
To simplify notation, we suppress the dependence of the statistical objects on the lag index $h$ and the time index  $t$. 

Throughout their paper, KP views nonnegative weights, $\omega(\cdot) \geq 0$, as the gold standard for an estimand to represent a ``meaningful causal summary''. Their justification is that if the region where $\omega(\cdot)$ is negative has a positive measure, then there exists a $g'(\cdot)$ that takes sufficiently large positive value therein to render $\beta$ negative even when $g^{\prime}(\cdot)$ is positive everywhere in $\mathcal{X}$. This claim is valid if one is fully agnostic about $g^{\prime}(\cdot)$. What is missing, however, is an assessment of the likelihood that negative weights lead to a reversal of the sign of $\beta$ in typical macroeconomic applications.

As KP note as motivation: “...both macroeconomic theorists and policymakers
think nonlinearities are important”. This statement reflects that macroeconomic
theorists and policymakers have some ideas about what are
admissible shapes and structures of the average marginal effect rather than being fully agnostic about $g'(x)$. Thus, even in cases where $\omega(x)$ can take negative values, there can be some credible restrictions available on the shape of $g'(x)$ under which the negative
weights are rendered inconsequential. Since we use macroeconomic theory to   guide empirical implementations in macroeconomics, we should also exploit it to assess whether the negative weight is a serious concern or not in a given context.


In this spirit, we show a proposition below to illustrate that shape restrictions on $g^{\prime}(\cdot)$ can resolve the issue of negative weight. In addition, these restrictions can even guarantee the existence of a representation $\beta = \int_{\mathcal{X}} \tilde{\omega}(x) g^{\prime}(x) dx$ with nonnegative weights $\tilde{\omega}(\cdot) \geq 0$, where $\tilde{\omega}(\cdot)$ is a transformation of the original potentially negative weights $\omega(\cdot)$.



To provide a formal statement, let us partition the domain of the shocks $\mathcal{X}$ into regions according to the sign of $\omega(\cdot)$:
\[
\mathcal{X}^{-}:=\{x:\omega(x)<0\},\;\mathcal{X}^{+}:=\{x:\omega(x)\geq0\}.
\]
Our proposition restricts $g'(x)$ to behave similarly in the negative weight region $\mathcal{X}^{-}$ and the positive weight region $\mathcal{X}^{+}$. Given a weight function \(\omega(x)\) and a marginal effect function \(g(x)\), we aim to find 
an injective link function $Q(\cdot)$, which is continuously first-order differentiable 
on $\mathcal{X}^-$ (except for a finite number of points) and maps each $x \in \mathcal{X}^{-}$ to some $Q(x) \in \mathcal{X}^{+}$ with an equal value of $g'(\cdot)$, i.e.,
\begin{equation}
g'(x) = g'(Q(x)). \label{A1}
\end{equation}
 In addition, we require the link function to have the property that the weight at every $x \in \mathcal{X}^{-}$ is dominated by the weight at $Q(x) \in \mathcal{X}^{+}$ in the sense that
\begin{equation}
\omega(x) + \omega(Q(x)) \geq 0. \label{A2}
\end{equation}
Availability of such link function 
$Q$ satisfying conditions (\ref{A1}) and (\ref{A2}) constrains the shape of $g'(x)$ in such a way that we can shift $g'(\cdot)$ on $\mathcal{X}^{-}$ to match $g'(\cdot)$ on $\mathcal{X}^{+}$ subject to the positive \textit{net} weight condition (\ref{A2}). 

Define $G(\mathcal{X}^{-}) = \{Q(x) : x \in \mathcal{X}^{-}\}$. Assuming that (\ref{A1}) and (\ref{A2}) hold, consider the transformed weights $\tilde{\omega}(\cdot)$ defined as follows:
\begin{equation}\label{checkweight}
\tilde{\omega}(x) = 
\begin{cases} 
\omega(Q^{-1}(x)) + \omega(x) & \text{for } x \in G(\mathcal{X}^{-}), \\
\omega(x) & \text{for $x \in \mathcal{X}^{+}\setminus G(\mathcal{X}^{-})$},
\\0 & \text{otherwise}.
\end{cases}
\end{equation}
These transformed weights $\tilde{\omega}(\cdot)$ are nonnegative by construction and supported on $\mathcal{X^{+}}$. Under additional conditions on the Jacobian of $Q(\cdot)$, the next proposition shows that we can express $\beta = \int_{\mathcal{X}} \omega(x) g'(x) dx$ as a weighted average of $g'(x)$ with positive weights $\tilde{\omega}(x)$.

\begin{proposition} \label{lem:link_Q}
 Suppose that a differentiable link function $Q: \mathcal{X}^{-} \to \mathcal{X}^{+}$ satisfies conditions (\ref{A1}), (\ref{A2}), and the derivative condition for $Q(\cdot)$,
\begin{equation}
  \int_{\mathcal{X}^{-}} [{\omega}(x)(1- Q'(x) )] g'(x) dx =0. \label{A3}
\end{equation}
Then, the following statements hold.
\begin{enumerate}
\item $\tilde{\omega}(\cdot) \geq 0$ constructed in (\ref{checkweight}) satisfies
\begin{equation} 
\int_{\mathcal{X}} \omega(x) g'(x) dx = \int_{\mathcal{X} \setminus \mathcal{X}^{-}} \tilde{\omega}(x) g'(x) dx. 
\end{equation}
\item If $Q(\cdot)$ additionally satisfies
\begin{equation}
\int_{\mathcal{X}^{-}} [\omega(x)(1-Q'(x))]dx=0, \label{A4}
\end{equation}
then $\int_{\mathcal{X}} \omega(x)  dx = \int_{\mathcal{X} \setminus \mathcal{X}^{-}} \tilde{\omega}(x)dx $ holds.
\end{enumerate}
\end{proposition}
See Appendix \ref{app:change_weight} for a proof. This proposition, which is new to the literature to our knowledge, shows that weighted-average estimands with negative weights $\omega(\cdot)$ can be salvaged by finding an alternative representation with positive weights $\tilde{\omega}(\cdot)$. The existence of $\tilde{\omega}(\cdot)$ relies on $g'(x)$ satisfying a set of shape conditions guaranteeing the existence of the link function characterized by (\ref{A1}), (\ref{A2}), and (\ref{A3}). With the additional condition (\ref{A4}) imposed, we can also ensure that the transformed weights $\tilde{\omega}(\cdot)$ sum to the same value as the original weights $\omega(\cdot)$. The following examples show how  this proposition can be applied.



\begin{figure}[h]
	\caption{\label{fig:example_1} Example \ref{exm:linear}}
	
	\centering
	
	\includegraphics[scale=0.45]{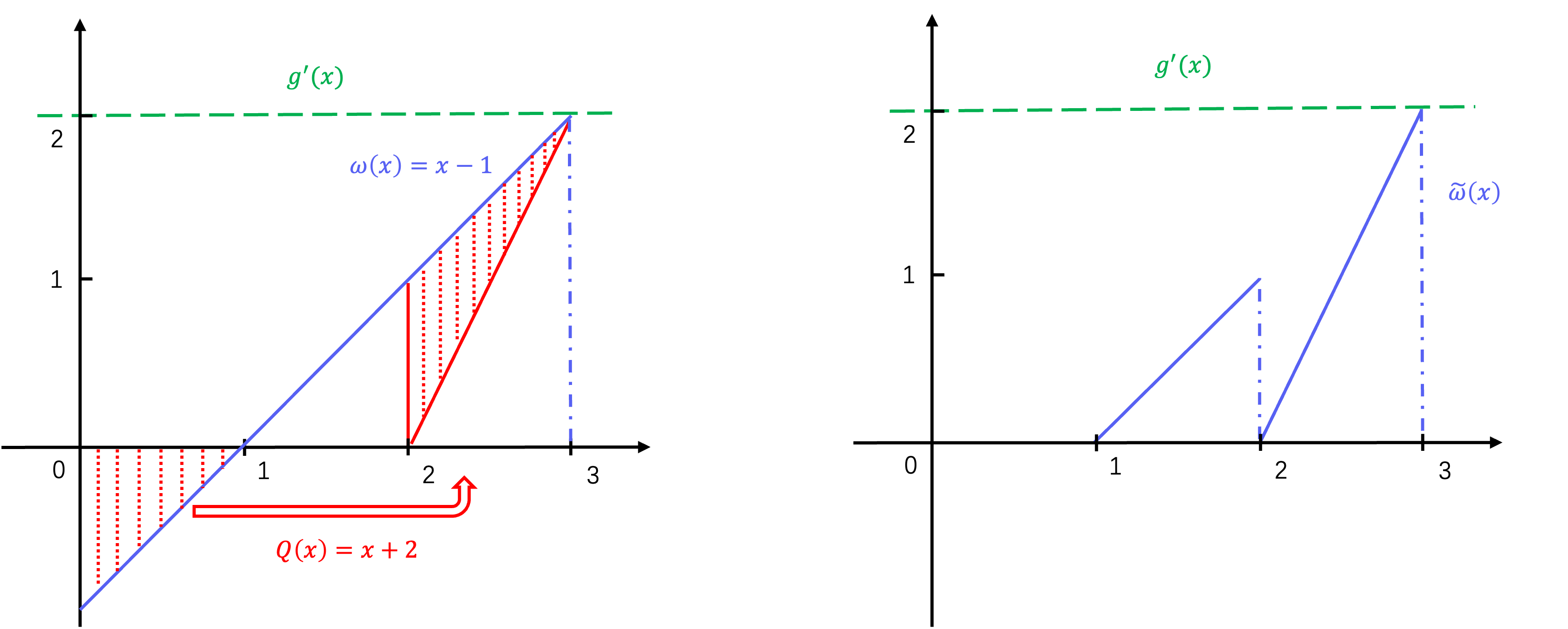}
		\noindent\begin{minipage}[t]{1\columnwidth}%
		
	\end{minipage}
\end{figure}

\begin{exm}\label{exm:linear}

A simple example is as follows: consider the weight function \(\omega(x) = x - 1\) and the marginal effect function \(g(x) = 2x - 3\). Define the compact support \(\mathcal{X} = [0, 3]\). This weight function \(\omega(x)\) assigns weights that are negative for \(x < 1\) and positive for \(x > 1\), splitting \(\mathcal{X}\) into \(\mathcal{X}^{-} = [0, 1)\) where \(\omega(x) < 0\) and \(\mathcal{X}^{+} = [1, 3]\) where \(\omega(x) \geq 0\). \(g(x)\) has a constant derivative \(g'(x) = 2\), which corresponds to the average marginal effect. In this case, we construct the link function \(Q(x) = x + 2\), which shifts \(x\) and has derivative \(Q'(x) = 1\); this maps \(\mathcal{X}^{-}\) to \(G(\mathcal{X}^{-}) = [2, 3) \subset \mathcal{X}^{+}\). The shift induces a new weight function:
\[
\tilde{\omega}(x) =
\begin{cases} 
\omega(x - c) + \omega(x) = (x - 2 - 1) + (x - 1) = 2x - 4 & \text{for } x \in G(\mathcal{X}^{-}) = [2, 3), \\
\omega(x) = x - 1 & \text{for } x \in \mathcal{X}^{+} \setminus G(\mathcal{X}^{-}) = [1, 2) \cup \{3\}, \\
0 & \text{otherwise in } [0, 3].
\end{cases}
\]
Now we check conditions  (\ref{A1}),  (\ref{A2}),  (\ref{A3}) and  (\ref{A4}). As \( g'(x) = 2 = g'(x + 2) \), $g(x)$ satisfies condition (\ref{A1}) universally across \( [0, 1) \). Conditions (\ref{A3}) and (\ref{A4}) are trivial as \( Q'(x) = 1 \): \( \int_{\mathcal{X}^{-}} \omega(x) (1 - 1) \, dg(x) = \int_{0}^{1} (x - 1) \cdot 0 \cdot 2 \, dx = 0 \) and \( \int_{\mathcal{X}^{-}} \omega(x) (1 - 1) \, dx = \int_{0}^{1} (x - 1) \cdot 0 \, dx = 0 \). Condition (\ref{A2}) \( \omega(x) + \omega(x + 2) = (x - 1) + (x + 2 - 1) = 2x  \geq 0 \) holds for \( x \in [0, 1) \) , and \( \tilde{\omega}(x) = 2x - 4 \geq 0 \) for \( x \in [2, 3) \) (where \( x \geq 2 \)), while \( \tilde{\omega}(x) = x - 1 \geq 0 \) for \( x \in [1, 2) \cup \{3\} \). 
 See Figure \ref{fig:example_1} for an illustration of Example \ref{exm:linear}.
\end{exm}

Our intention with Proposition \ref{lem:link_Q} is to highlight an unexplored connection between restrictions on $g'(\cdot)$ and the harmlessness of negative weights. We do not argue that the conditions of this proposition are likely to be met in all common macroeconometric applications, but further research on salvaging negative weight estimands through restrictions on $g'(\cdot)$ may be an interesting direction for future research following the contributions of KP.

To simplify exposition, Proposition \ref{lem:link_Q} assumes the existence of a differentiable link function $Q(\cdot)$. We can obtain a more general version of this proposition, which does not involve a link function, by imposing a dominance condition on the measures induced by $\omega(\cdot)$ and the mappings between $g^{\prime}(\cdot)$ and $\mathcal{X}^{-}$ and $\mathcal{X}^{+}$. See Proposition \ref{lem:gen} in Appendix \ref{app:a general approach} for more details. Additionally, in Example \ref{example_KP_P2} of Appendix \ref{app:a general approach}, we also revisit KP's Proposition 2 to show our approach can restore the interpretation of
the linear regression coefficient as an average-effect estimator with positive weights when the shock is positive.




\section{Comment 2: Time Heterogeneity of Causal Effects}

One of the fundamental assumptions underlying the analysis of KP is that, while the average structural equation can depend on the time lag between the period in which the outcome is observed and the period in which treatment is given, it does not depend on the time index $t$. That is, the causal effect is assumed to be stationary. However, in reality, nonstationary causal effects can be as great a concern as nonlinear structural equations. 

For a binary $X_t$ which satisfies sequential unconfoundedness and has known propensity scores, \cite{bojinov2019time} shows that the regression coefficient of $Y_t$ on $X_t \in \{0, 1\}$ estimated with inverse propensity score weighting (i.e., the difference in inverse-propensity-score weighted means between $X_t = 1$ and $X_t = 0$ observations) provides an unbiased estimate of

\[
\frac{1}{T} \sum_{t=1}^T \mathbb{E}[\psi_t(1, U_t) - \psi_t(0, U_t) \mid \mathcal{F}_{t-1}], \quad \mathcal{F}_{t-1} = \sigma(Y_{t-1}, X_{t-1}, Y_{t-2}, X_{t-2}, \dots),
\]
where $\mathcal{F}_{t-1}$ denotes the set of conditioning variables (information set) consisting of observable variables up to $t-1$. Each summand can be viewed as the conditional average treatment effect at $t$ given the history $\mathcal{F}_{t-1}$, with the structural equation $\psi_t(\cdot, \cdot)$ allowed to vary in $t$. In turn, the whole estimand can be viewed as a sample average of conditional average effects over the sampled periods. \cite{casini2024identification} shows a similar result when $X_t$ is continuous in a high-frequency context with infill asymptotics. 

In the case of a continuous $X_t$, can the regression coefficient $\hat{\beta}$ be interpreted as an average of causal effects even if the structural equation is nonstationary? Can we allow instability in the economic environment to the extent that the time series of observables have unit roots? Can we improve the statistical precision of the estimator or the interpretability of the estimand by finding alternative methods that place greater weight on more recent observations?
Answering these questions would help clarify whether the unrestricted heterogeneity of causal effects allowed in the cross-sectional setting can be extended to the time-series setting. To our knowledge, little is known about these questions beyond the binary case.



\section{Comment 3: Make ``The Good'' Great?}
KP argues that the average marginal effect identified by $\beta$ is an estimand of interest as it can be estimated well even when the sample size is limited, as is common in empirical macroeconomic research. This argument is appealing and shares a common motivation with the procedure of selecting estimands based on their efficient asymptotic variances, as proposed in \citep{Crump_etal_2009} for the case of treatment effect estimation in microeconometrics. 

A counterargument is that, instead of narrowing the choice of estimand based on identifiability or estimation precision, one should choose the estimand of interest based on the ultimate goal of analysis and the type of policy questions that researchers wish to investigate. 
This raises the question of how $\beta$ can be connected to counterfactual policy analysis and ex ante policy decisions.
However, given that the analytical expression of the weights in $\beta$ is driven purely by the characteristics of the data generating processes, we doubt the existence of a direct link between $\beta$ and a planner's policy decision problem.  



Making $\beta$ policy-relevant would turn ``the good'' into ``the great''. One way to achieve this is if, based on economic theory or available background knowledge, we can justify the assumption that the sign or magnitude of $\beta$ is informative for current or future optimal policy. 
Building upon the potential outcome time-series framework \citep{angrist2018semiparametric, bojinov2019time, rambachan2021when},  \cite{kitagawa2022policy} formulates a policy choice problem in the time-series setting with a binary $X_t$ and studies ex ante policy choices while accounting for nonstationary causal effects. Our approach relies on the ``invariance of welfare ordering''. This assumption states that the ex post welfare ranking of past policies provides reliable information about the ex ante welfare ranking that is relevant to current policy decisions. For the case of binary $X_t$, a sufficient condition for the welfare ordering to be invariant is that the sign of $\beta$ coincides with the sign of the welfare impact of the policy if it were implemented today. 


Extending this type of approach to the case of a continuous $X_t$ is an area that remains unexplored. Establishing how an estimand involving negative weights in its weighted-average interpretation interacts with the validity of the invariance of welfare ordering or a similar assumption could be an insightful exercise. 



\section{Conclusion}

KP have performed a thorough and insightful reverse econometric analysis of the estimands that are commonly used in empirical macroeconomics. We have presented sketch ideas on how the bad can be salvaged (Comment 1) and how the good can be made great (Comment 3), but leave further development and assessment of the usefulness of these ideas to future research. 

We want to highlight that KP contributes to econometrics more broadly by bridging microeconometrics and macroeconometrics, highlighting the potential for greater synergies between these two areas and paving the way for future research that benefits from the strengths of both micro and macroeconometric analysis.

\newpage
\appendix

\section*{Appendix}
\section{Additional example of Proposition \ref{lem:link_Q}}
\begin{exm}

As a second example, consider the weight function \(\omega(x) = x - 1\) and a marginal effect function \(g(x)\), defined over the compact support \(\mathcal{X} = [0, 2]\). This weight function \(\omega(x)\) assigns negative weights for \(x < 1\) and positive weights for \(x > 1\), partitioning \(\mathcal{X}\) into the negative weight region \(\mathcal{X}^{-} = [0, 1]\) where \(\omega(x) < 0\) and the positive weight region \(\mathcal{X}^{+} = [1, 2]\) where \(\omega(x) \geq 0\). The derivative of the marginal effect function is specified piecewise as \(g'(x) = 2 - x + 12x^2 - 12x^3\) for \(x \in [0, 1]\) and \(g'(x) = x\) for \(x \in [1, 2]\). Hence, \(g(x) = 2x - \frac{1}{2}x^2 + 4x^3 - 3x^4\) for \(x \in [0, 1]\) and \(g(x) = \frac{1}{2}x^2 + 2\) for \(x \in [1, 2]\). In this case, we use the link function \(Q(x) = 2 - x + 12x^2 - 12x^3\), which has derivative \(Q'(x) = -1 + 24x - 36x^2\). This maps \(\mathcal{X}^{-}\) to \(G(\mathcal{X}^{-}) = [1, 2]\), and its inverse \(Q^{-1}(x)\) is a root of the cubic equation \(12y^3 - 12y^2 + y - x + 2 = 0\). This shift induces a transformed weight function \(\tilde{\omega}(x)\), given by 
\[
\tilde{\omega}(x) = 
\begin{cases} 
(x - 1) + (Q^{-1}(x) - 1) & \text{for } x \in [1, 2], \\
0 & \text{otherwise in } [0, 2].
\end{cases}
\]
This setup satisfies the required conditions: Condition (\ref{A1}) holds since \(g'(x) = g'(Q(x)) = 2 - x + 12x^2 - 12x^3\) for \(x \in \mathcal{X}^{-}\), Condition (\ref{A2}) is met as \(\omega(x) + \omega(Q(x)) = 12x^2(1 - x) \geq 0\) for \(x \in [0, 1]\), Condition (\ref{A3}) is verified by \(\int_0^1 (x - 1)(1 - Q'(x)) Q(x) \, dx = 0\), and Condition (\ref{A4}) holds with \(\int_0^1 (x - 1)(1 - Q'(x)) \, dx = 0\).

\end{exm}
\section{A generalization of Proposition \ref{lem:link_Q}}\label{app:a general approach}

Proposition \ref{lem:link_Q}  requires the existence and identification of a link function $Q(\cdot)$. This requirement can be relaxed to a measure dominance condition. To this end,  define
$Y\equiv g^{\prime}(X).$  and  let $\mathcal{Y}\subseteq\mathbb{R}$
denote the support of random variable $Y$.

Consider two set-valued mappings from a measurable set in $\mathcal{Y}$ to a measurable set in $\mathcal{X}\cup\emptyset$:
For any measurable set $B\subseteq\mathcal{Y}$,
\begin{align}
g_{-}^{-1}(B) & =\{x:x\in\left(g^{\prime}\right)^{-1}(B)\cap\mathcal{X}^{-}\},\nonumber \\
g_{+}^{-1}(B) & =\{x:x\in\left(g^{\prime}\right)^{-1}(B)\cap\mathcal{X}^{+}\}.\label{eq:induced_mu}
\end{align}
Note that in these definitions, we do not assume $g'(\cdot)$ is an invertible function and $ \left(g^{\prime}\right)^{-1}(x)$ can be set-valued. For example, if $g^{\prime}(x)=x^{2},$ we have $\left(g^{\prime}\right)^{-1}(\{1\})=\{1,-1\}$.

Furthermore, define 
\begin{equation*}
\mu_{y^{-}}(B)  =-\int_{x\in g_{-}^{-1}(B)}\omega(x)dx,\quad
\mu_{y^{+}}(B)  =\int_{x\in g_{+}^{-1}(B)}\omega(x)dx,
\end{equation*}
 to be the measures on $\mathcal{Y}$ induced by the weight function $\omega(\cdot)$,
and the mappings $g_{-}^{-1}$ and $g_{+}^{-1}$. 

The required measure dominance condition
is: for any measurable set $B\subseteq\mathcal{Y}$, it holds that
\begin{equation}
\mu_{y^{-}}(B)\leq\mu_{y^{+}}(B)\label{eq:measure dominance}.
\end{equation}

Intuitively, this condition ensures that each negative
weight can be covered by a positive weight with the same value
of treatment effect, $g^{\prime}(x)$. Since condition \eqref{eq:measure dominance}
is defined with respect to any measurable subset of $\mathcal{Y}$,
it is general and does not explicitly impose any specific functional
form on $\omega(\cdot)$ and $g^{\prime}(\cdot)$, and the link function $Q(\cdot)$ is no longer required.

In addition, this condition is similar to yet stronger
than the standard definition of measure dominance (or absolute continuity;
see p. 422 \citeauthor{billingsley1995}, \citeyear{billingsley1995}). We need not only that $\mu_{y^{-}}(B)=0\Rightarrow\mu_{y^{+}}(B)=0$,
but also that $\mu_{y^{-}}(B)\leq\mu_{y^{+}}(B)$ for all measurable sets
within $\mathcal{Y}$. Note that the measures $\mu_{y^{-}}$ and $\mu_{y^{+}}$
are defined as general measures rather than probability measures;
we do not require that $\int_{\mathcal{Y}}d\mu_{y^{\pm}}(y)=1$.

Define the subset of $\mathcal{X}^{+}$ whose members possess
matched points in $\mathcal{X}^{-}$ to be
\begin{equation}
\mathcal{X}_{m}^{+}=\{x\in\mathcal{X}^{+}:\exists\;z\in\mathcal{X}^{-},\text{such that }g^{\prime}(z)=g^{\prime}(x)\}.\label{eq:X_positive_matched}
\end{equation}
Let $\mathcal{B}(x,r)$ denote an Euclidean ball centered around $x$ with a radius of $r$.
Then, the transformed weights are given by:
\begin{equation}
\tilde{\omega}(x)=\begin{cases}
\omega(x) & x\in\mathcal{X}^{+}\backslash\mathcal{X}_{m}^{+}\\
\lim_{r\to 0}\frac{\int_{v\in g_{+}^{-1}(\left\{ g^{\prime}(\mathcal{B}(x,r))\right\} )}\omega(v)dv+\int_{v\in g_{-}^{-1}(\left\{ g^{\prime}(\mathcal{B}(x,r))\right\} )}\omega(v)dv}{\int_{v\in g_{+}^{-1}(\left\{ g^{\prime}(\mathcal{B}(x,r))\right\} )}dv} & x\in\mathcal{X}_{m}^{+}\\
0 & x\in\mathcal{X}^{-}
\end{cases}.\label{eq:T_weight}
\end{equation}

\begin{lemma} \label{lem:gen} Assume that  $\mathcal{X}$ is compact. Let both $\omega(\cdot)$ and $g^{\prime}(\cdot)$ be continuous functions defined on $\mathcal{X}$. 
Under condition \eqref{eq:measure dominance}, the
transformed weight function $\tilde{\omega}(\cdot)$ has the following
properties:
\begin{enumerate}
\item $\int\omega(x)g^{\prime}(x)dx=\int\tilde{\omega}(x)g^{\prime}(x)dx$;
\item $\tilde{\omega}(x)\geq0$ for all $x\in\mathcal{X}$.
\end{enumerate}
\end{lemma}

\begin{exm}[Constant marginal effect]
Assume that the marginal effect function, $g^{\prime}(\cdot)$ is a constant function,
i.e., $g^{\prime}(X)\equiv y$. Then, we have $\left(g^{\prime}\right)^{-1}(\{y\})=\mathcal{X}$.
Furthermore,
\begin{equation*}
g_{-}^{-1}(\{y\})  =\mathcal{X}^{-},\quad
g_{+}^{-1}(\{y\})  =\mathcal{X}^{+}.
\end{equation*}
The measure dominance condition \eqref{eq:measure dominance} can thus be written as
$
-\int_{x\in\mathcal{X}^{-}}\omega(x)dx\le\int_{x\in\mathcal{X}^{+}}\omega(x)dx
$, i.e., \begin{equation}\int_{x\in\mathcal{X}^{+}}\omega(x)dx+\int_{x\in\mathcal{X}^{-}}\omega(x)dx=\int_{x\in\mathcal{X}}\omega(x)dx\geq0. \label{eq:const_gen}\end{equation}
If this weight function satisfies 
$\int_{x\in\mathcal{X}}\omega(x)dx=1$,
Condition \eqref{eq:measure dominance}
is always satisfied. By the definition of $\mathcal{X}^{+}$ and $\mathcal{X}^{-}$, a further implication of $\int_{x\in\mathcal{X}}\omega(x)dx=1$ is that $\int_{v\in\mathcal{X}^{+}}dv>0$.
The transformed weight function \eqref{eq:T_weight} is:
\[
\tilde{\omega}(x)=\begin{cases}
\frac{1}{\int_{v\in\mathcal{X}^{+}}dv} & x\in\mathcal{X}^{+}\\
0 & x\in\mathcal{X}^{-}
\end{cases}.
\]
\end{exm}

\begin{exm} \label{example_KP_P2} [Proposition 2 in KP revisited]

Proposition 2 of KP considers a regression with an additional
quadratic term, and shows that in this case negative weights are generally unavoidable. Here we show that under certain assumptions our proposed approach enables us to recover positive weights.

Under the assumptions of KP's Proposition 2,
equation (12) of KP shows that for any $z\in\mathbb{R}$, 
\begin{align*}
\bar{\beta}(z) & =\mathbb{E}\left[\left(1+Xz\right)g^{\prime}(X)\right]=\int\left(1+z\cdot x\right)g^{\prime}(x)\phi(x)dx,
\end{align*}
where $\phi(\cdot)$ is the density function of the standard normal distribution. Then, for
a given $z$, the weight function takes the form \begin{equation}\label{eq:w_P2_KP}
\omega(x)=\left(1+z\cdot x\right)\phi(x).
\end{equation}
Assume that:
\[
\text{The marginal effect function, \ensuremath{g^{\prime}}(\ensuremath{\cdot}), is symmetric about zero.}
\]
\noindent \emph{Case 1: $z\geq0$, a positive shock.}

The weight function \eqref{eq:w_P2_KP} is negative for $x<-\frac{1}{z}$. Thus, 
$\mathcal{X}=\mathbb{R}$, $\mathcal{X}^{-}=[-\infty,-\frac{1}{z})$,
and $\mathcal{X}^{+}=[-\frac{1}{z},\infty)$. By the
symmetry of $g^{\prime}(\cdot)$, we have $\mathcal{X}_{m}^{+}=[\frac{1}{z},\infty)$.
Since both $g^{\prime}(\cdot)$ and $\phi(\cdot)$ are symmetric,
the inequality
\begin{align*}
-\omega(-x) & =-\left(1-z\cdot x\right)g^{\prime}(-x)\phi(-x) =\left|1-z\cdot x\right|g^{\prime}(x)\phi(x)\\
 & \leq\max\left(1,\left|z\cdot x\right|\right)g^{\prime}(x)\phi(x)
  \leq\left(1+z\cdot x\right)g^{\prime}(x)\phi(x)=\omega(x)
\end{align*}
holds pointwise for every $z>0$ and $x\in\mathcal{X}_{m}^{+}\equiv[\frac{1}{z},\infty)$.
Note that if $x\in\mathcal{X}_{m}^{+}\equiv[\frac{1}{z},\infty)$,
then $-x\in[-\infty,-\frac{1}{z})\equiv\mathcal{X}^{-}$. Consequently,
condition \eqref{eq:measure dominance} is satisfied.

Then, the second row of the weight function \eqref{eq:T_weight} becomes
\begin{align*}
\left[\left(1-z\cdot x\right)\phi(-x)+\left(1+z\cdot x\right)\phi(x)\right]\frac{dx}{dx} & =\left[\left(1-z\cdot x\right)+\left(1+z\cdot x\right)\right]\phi(x)
  =2\phi(x).
\end{align*}
The transformed weight function can be written as:
\[
\tilde{\omega}(x)=\begin{cases}
\left(1+z\cdot x\right)\phi(x) & x\in[-1/z,1/z]\\
2\phi(x) & x\in[\frac{1}{z},\infty)\\
0 & x\in[-\infty,-\frac{1}{z})
\end{cases}.
\]
 This weight function is always non-negative, and $\int\omega(x)g^{\prime}(x)dx=\int\tilde{\omega}(x)g^{\prime}(x)dx$
by construction.\\

\noindent \emph{Case 2: $z<0$, a negative shock}

In this case condition \eqref{eq:measure dominance} is generally
not satisfied. Note that $[1/z,-1/z)\in\mathcal{X}^{-}$. However, in general, there
is no match in $\mathcal{X}^{+}$ for an $x\in[1/z,-1/z)$.

In summary, our approach can restore the interpretation of
the linear regression coefficient as an average-effect estimator with positive weights in the setup of Proposition 2 of
KP when the shock is positive.
\end{exm}
\section{Proofs}
\subsection{Proof of Proposition \ref{lem:link_Q}}\label{app:change_weight}
\begin{proof}
Recall that the first statement of the proposition is:

\[
\int_{\mathcal{X} \setminus \mathcal{X}^-}  \tilde{\omega}(x) g'(x) dx = \int_{\mathcal{X}} {\omega}(x) g'(x) dx.
\]
The above statement is equivalent to,
\[
\int_{\mathcal{X} \setminus \mathcal{X}^{-}}  [\omega(x) -\tilde{\omega}(x) ]g'(x) dx + \int_{\mathcal{X}^{-}} {\omega}(x) g'(x) dx =0.
\]
Thus by the definition of $\tilde{\omega}(x)$,
\[
-\int_{G(\mathcal{X}^{-})}  [\omega(Q^{-1}(x)) ]g'(x) dx + \int_{\mathcal{X}^{-}} {\omega}(x) g'(x) dx =0.
\]
By condition (\ref{A1}), we have 

\[
-\int_{G(\mathcal{X}^{-})}  [\omega(Q^{-1}(x)) ]g'(Q^{-1}(x)) dx + \int_{\mathcal{X}^{-}} {\omega}(x) g'(x) dx =0.
\]
Take $v = Q^{-1}(x)$, we can write 
\[
-\int_{\mathcal{X}^{-}}  \omega(v) g'(v) dQ(v) + \int_{\mathcal{X}^{-}} {\omega}(x) g'(x) dx =0.
\]
This requires that
\[
-\int_{\mathcal{X}^{-}}  \omega(v) Q'(v) g'(v) dv + \int_{\mathcal{X}^{-}} {\omega}(x) g'(x) dx =0.
\]
Thus, we have, 
\[
\int_{\mathcal{X}^{-}} \{{\omega}(x)- \omega(x) Q'(x) \} g'(x) dx =0.
\]
This leads to condition (\ref{A3})
\[
\int_{\mathcal{X}^{-}} \{{\omega}(x)- \omega(x) Q'(x) \}  d g(x) =0.
\]

Next, to prove the second statement of Proposition \ref{lem:link_Q},
we need to show that:
$$\int_{\mathcal{X}} \omega(x)  dx = \int_{\mathcal{X} \setminus \mathcal{X}^{-}} \tilde{\omega}(x)dx.$$

To prove this, we have 
\[\int_{\mathcal{X} \setminus \mathcal{X}^{-}} \tilde{\omega}(x)dx=\int_{G(\mathcal{X}^{-})} [\omega(Q^{-1}(x)) + \omega(x)  ]dx+\int_{\mathcal{X} \setminus (\mathcal{X}^{-} \cup G(\mathcal{X}^{-}))} [ \omega(x)  ]dx
\]

\[\int_{\mathcal{X} \setminus \mathcal{X}^{-}} \tilde{\omega}(x)dx=\int_{G(\mathcal{X}^{-})} [\omega(Q^{-1}(x)) + \omega(x)  ]dx+\int_{\mathcal{X} \setminus (\mathcal{X}^{-} \cup G(\mathcal{X}^{-}))} [ \omega(x)  ]dx
\]

\[\int_{\mathcal{X} \setminus \mathcal{X}^{-}} \tilde{\omega}(x)dx=\int_{\mathcal{X}^{-}} [\omega(x)Q'(x)  ]dx+\int_{G(\mathcal{X}^{-})} [\omega(x)]dx+\int_{\mathcal{X} \setminus (\mathcal{X}^{-} \cup G(\mathcal{X}^{-}))} [ \omega(x)  ]dx
\]
This leads to the following condition:
\[\int_{\mathcal{X}^{-}} [\omega(x)(1-Q'(x))]dx=0
\]

Thus, the two conditions we need to ensure statements 1 and 2 of Proposition \ref{lem:link_Q} are, respectively:

\[\int_{\mathcal{X}^{-}} [\omega(x)(1-Q'(x))]dx=0
\]

\[
\int_{\mathcal{X}^{-}} \{{\omega}(x)- \omega(x) Q'(x) \}  d g(x) =0.
\]

\end{proof}

\subsection{Proof of Proposition \ref{lem:gen}} \label{proof:lem2}

\begin{proof}
We prove this proposition in three steps.
	\begin{itemize}
		\item[Step 1.] We show that the weight function $\tilde{\omega}(\cdot)$ is well defined.
		\item[Step 2.] We show that the two statements of the proposition hold for a sequence of effect functions,  $\{g_n(\cdot)\}_{n\in\mathbb{N}}$, and the following weight function
		\begin{equation}
			\tilde{\omega}_n(x)=\begin{cases}
				\omega(x) & x\in\mathcal{X}^{+}\backslash\mathcal{X}_{m}^{+}\\
				\frac{\int_{v\in g_{+}^{-1}(\left\{ g_n^{\prime}(x)\right\} )}\omega(v)dv+\int_{v\in g_{-}^{-1}(\left\{ g_n^{\prime}(x)\right\} )}\omega(v)dv}{\int_{v\in g_{+}^{-1}(\left\{ g_n^{\prime}(x)\right\} )}dv} & x\in\mathcal{X}_{m}^{+}\\
				0 & x\in\mathcal{X}^{-}
			\end{cases},\label{eq:step_w}
		\end{equation}
		where $g_n(\cdot)$ is a undersmoothed version of $g(\cdot)$, such that $g_n^{\prime}(x)$ takes finitely many values on $\mathcal{X}$, and $\lim_{n\to \infty }g_n^{\prime}(x)=g(x) $ for all $x\in \mathcal{X}$. In Step 2, $g_+^{-1}(\cdot)$, $g_-^{-1}(\cdot)$, and $\mathcal{X}^{+}_m$ are defined with respect to $g_n(\cdot)$.
		\item [Step 3.]  We conclude the proof by employing Lebesgue's Dominated Convergence Theorem.
	\end{itemize}
	\begin{rem}
	 If $g^\prime(\cdot)$ is discrete and takes finitely many values, only Step 2 is required to complete the proof. The assumption that $g^\prime(\cdot)$ is continuous is made to simplify the discussion of Steps 1 and 3. This assumption is generally not required for the proposition.
	\end{rem}
\noindent \emph{Step 1.}
 
For all $x\in\mathcal{X}_{m}^{+}$, we have \begin{align}
0\leq\frac{\int_{v\in g_{+}^{-1}(\left\{ g^{\prime}(\mathcal{B}(x,r))\right\} )}\omega(v)dv+\int_{v\in g_{-}^{-1}(\left\{ g^{\prime}(\mathcal{B}(x,r))\right\} )}\omega(v)dv}{\int_{v\in g_{+}^{-1}(\left\{ g^{\prime}(\mathcal{B}(x,r))\right\} )}dv} \leq \frac{\int_{v\in g_{+}^{-1}(\left\{ g^{\prime}(\mathcal{B}(x,r))\right\} )}\omega(v)dv}{\int_{v\in g_{+}^{-1}(\left\{ g^{\prime}(\mathcal{B}(x,r))\right\} )}dv}\leq \sup_{x\in\mathcal{X}} \omega(x).\label{eq:P2_step1}
\end{align}
By the continuity of $g^{\prime}(\cdot)$, $\int_{v\in g_{+}^{-1}(\left\{ g^{\prime}(\mathcal{B}(x,r))\right\} )}dv$ is strictly positive if $r>0$. The first inequality therefore follows from Condition \eqref{eq:measure dominance}. 
The second inequality follows as, by the definitions of $g_+^{-1}$ and $g_-^{-1}$ in \eqref{eq:induced_mu}, $\int_{v\in g_{-}^{-1}(\left\{ g^{\prime}(\mathcal{B}(x,r))\right\} )}\omega(v)dv$ is negative, and $\int_{v\in g_{+}^{-1}(\left\{ g^{\prime}(\mathcal{B}(x,r))\right\} )}\omega(v)dv$ is positive. The existence of $\sup_{x\in\mathcal{X}} \omega(x)$ is implied by the continuity of $\omega(\cdot)$ and the compactness of $\mathcal{X}$. Note that under Condition \eqref{eq:measure dominance},  \eqref{eq:P2_step1} holds for all $r>0$. Consequently, $0\leq\tilde{\omega}(x)\leq\sup_{x\in\mathcal{X}} \omega(x)<\infty $. 

\noindent \emph{Step 2.}

First, we show the equivalence result,  $\int\omega(x)g_n^{\prime}(x)dx=\int\tilde{\omega}_n(x)g_n^{\prime}(x)dx$. Note that
\begin{align*}
\int\omega(x)g_n^{\prime}(x)dx & =\int_{x\in\mathcal{X}^{+}\backslash\mathcal{X}_{m}^{+}}\omega(x)g_n^{\prime}(x)dx+\int_{x\in\mathcal{X}_{m}^{+}}\omega(x)g_n^{\prime}(x)dx +\int_{x\in\mathcal{X}^{-}}\omega(x)g_n^{\prime}(x)dx,
\end{align*}
and
\begin{align*}
\int\tilde{\omega}_n(x)g_n^{\prime}(x)dx & =\int_{x\in\mathcal{X}^{+}\backslash\mathcal{X}_{m}^{+}}\tilde{\omega}_n(x)g_n^{\prime}(x)dx+\int_{x\in\mathcal{X}_{m}^{+}}\tilde{\omega}_n(x)g_n^{\prime}(x)dx +\int_{x\in\mathcal{X}^{-}}\tilde{\omega}_n(x)g_n^{\prime}(x)dx\\
 & =\int_{x\in\mathcal{X}^{+}\backslash\mathcal{X}_{m}^{+}}\omega(x)g_n^{\prime}(x)dx+\int_{x\in\mathcal{X}_{m}^{+}}\tilde{\omega}_n(x)g_n^{\prime}(x)dx+0,
\end{align*}
where the last equality follows by \eqref{eq:step_w}. To proceed, we must verify that
\begin{equation}
\int_{x\in\mathcal{X}_{m}^{+}}\tilde{\omega}_n(x)g_n^{\prime}(x)dx=\int_{x\in\mathcal{X}_{m}^{+}}\omega(x)g_n^{\prime}(x)dx+\int_{x\in\mathcal{X}^{-}}\omega(x)g_n^{\prime}(x)dx,\label{eq: middle_step_equiv}
\end{equation}
i.e., we want to show
\begin{align}
\int_{\mathcal{X}_{m}^{+}}\frac{\int_{v\in g_{+}^{-1}(\left\{ g_n^{\prime}(x)\right\} )}\omega(v)dv}{\int_{v\in g_{+}^{-1}(\left\{ g_n^{\prime}(x)\right\} )}dv}g_n^{\prime}(x)dx&=\int_{\mathcal{X}_{m}^{+}}\omega(x)g_n^{\prime}(x)dx,\\
\int_{\mathcal{X}_{m}^{+}}\frac{\int_{v\in g_{-}^{-1}(\left\{ g_n^{\prime}(x)\right\} )}\omega(v)dv}{\int_{v\in g_{+}^{-1}(\left\{ g_n^{\prime}(x)\right\} )}dv}g_n^{\prime}(x)dx&=\int_{\mathcal{X}^{-}}\omega(x)g_n^{\prime}(x)dx.
\end{align}
To this end, let us define
\begin{equation}
\mathcal{Y}^{-}:=\{y:g_{-}^{-1}(\{y\})\ne\emptyset\}.\label{eq:Y_minus}
\end{equation}
Recall that within the setup of Step 2, $g_+^{-1}(\cdot)$, $g_-^{-1}(\cdot)$, and $\mathcal{X}^{+}_m$ are defined with respect to $g_n(\cdot)$. Thus, $\mathcal{Y}$ is a finite set by the definition of $g_n(\cdot)$, and so is $\mathcal{Y}^-$. Then, based on the definitions
of $g_{+}^{-1}(\cdot)$ and $\mathcal{X}_{m}^{+}$, and the finiteness of $\mathcal{Y^{-}}$,
we can define the subregions of the integral of $x$.
For any (measurable) function $\delta(x)$, it holds that 
\begin{align}
\int_{x\in\mathcal{X}_{m}^{+}}\delta(x)dx & =\sum_{y\in\mathcal{Y^{-}}}\int_{x\in g_{+}^{-1}(\left\{ y\right\} )}\delta(x)dx,\label{eq: equal_two_regions}
\end{align}
i.e., for any $x\in\mathcal{X}_{m}^{+}\subseteq\mathcal{X}^{+}$,
the corresponding function value, $g_n^{\prime}(x)$, is always included in the set
of $y\in\mathcal{Y^{-}}.$ 
As a result,
\begin{align*}
\int_{x\in\mathcal{X}_{m}^{+}}\frac{\int_{v\in g_{+}^{-1}(\left\{ g_n^{\prime}(x)\right\} )}\omega(v)dv}{\int_{v\in g_{+}^{-1}(\left\{ g_n^{\prime}(x)\right\} )}dv}g_n^{\prime}(x)dx &
=\int_{x\in\mathcal{X}_{m}^{+}}\frac{\int_{v\in g_{+}^{-1}(\left\{ g_n^{\prime}(x)\right\} )}\omega(v)g_n^{\prime}(x)dv}{\int_{v\in g_{+}^{-1}(\left\{ g_n^{\prime}(x)\right\} )}dv}dx\\
&=\int_{x\in\mathcal{X}_{m}^{+}}\frac{\int_{v\in g_{+}^{-1}(\left\{ g_n^{\prime}(x)\right\} )}\omega(v)g_n^{\prime}(v)dv}{\int_{v\in g_{+}^{-1}(\left\{ g_n^{\prime}(x)\right\} )}dv}dx\\
&=\sum_{y\in\mathcal{Y^{-}}}\int_{x\in g_{+}^{-1}(\left\{ y\right\} )}\frac{\int_{v\in g_{+}^{-1}(\left\{ y\right\} )}\omega(v)g_n^{\prime}(v)dv}{\int_{v\in g_{+}^{-1}(\left\{ y\right\} )}dv}dx\nonumber \\
&=\sum_{y\in\mathcal{Y^{-}}}\int_{v\in g_{+}^{-1}(\left\{ y\right\} )}\omega(v)g_n^{\prime}(v)dv=\int_{x\in\mathcal{X}_{m}^{+}}\omega(x)g_n^{\prime}(x)dx,
\end{align*}
where the second equality follows as $g_n^{\prime}(v)=g_n^{\prime}(x)$ holds for any $v\in g_{+}^{-1}(\left\{ g_n^{\prime}(x)\right\} )$, the third equality
follows from \eqref{eq: equal_two_regions}, the fourth equality follows
from taking $\int_{v\in g_{+}^{-1}(\left\{ y\right\} )}\omega(v)g_n^{\prime}(v)dv$
out from the integral of $\int_{x\in g_{+}^{-1}(\left\{ y\right\} )}...dx$
and $\int_{x\in g_{+}^{-1}(\left\{ y\right\} )}\frac{1}{\int_{v\in g_{+}^{-1}(\left\{ y\right\} )}dv}dx=\frac{\int_{x\in g_{+}^{-1}(\left\{ y\right\} )}dx}{\int_{v\in g_{+}^{-1}(\left\{ y\right\} )}dv}=1$, and the last equality follows from \eqref{eq: equal_two_regions}.


Then, 
\begin{align}
\int_{x\in\mathcal{X}_{m}^{+}}\frac{\int_{v\in g_{-}^{-1}(\left\{ g_n^{\prime}(x)\right\} )}\omega(v)dv}{\int_{v\in g_{+}^{-1}(\left\{ g_n^{\prime}(x)\right\} )}dv}g_n^{\prime}(x)dx 
&=\int_{x\in\mathcal{X}_{m}^{+}}\frac{\int_{v\in g_{-}^{-1}(\left\{ g_n^{\prime}(x)\right\} )}\omega(v)g_n^{\prime}(x)dv}{\int_{\in g_{+}^{-1}(\left\{ g_n^{\prime}(x)\right\} )}dv}dx\nonumber \\
&=\int_{x\in\mathcal{X}_{m}^{+}}\frac{\int_{v\in g_{-}^{-1}(\left\{ g_n^{\prime}(x)\right\} )}\omega(v)g_n^{\prime}(v)dv}{\int_{\in g_{+}^{-1}(\left\{ g_n^{\prime}(x)\right\} )}dv}dx\nonumber \\
 & =\sum_{y\in\mathcal{Y^{-}}}\int_{x\in g_{+}^{-1}(\left\{ y\right\} )}\frac{\int_{v\in g_{-}^{-1}(\left\{ y\right\} )}\omega(v)g_n^{\prime}(v)dv}{\int_{v\in g_{+}^{-1}(\left\{ y\right\} )}dv}dx=\sum_{y\in\mathcal{Y^{-}}}\int_{v\in g_{-}^{-1}(\left\{ y\right\} )}\omega(v)g_n^{\prime}(v)dv,\label{eq:tilde_w_2}
\end{align}
where the second equality follows as $g_n^{\prime}(v)=g_n^{\prime}(x)$ holds for any $v\in g_{-}^{-1}(\left\{ g_n^{\prime}(x)\right\} )$, the third equality
follows from \eqref{eq: equal_two_regions}, and the last equality follows
from taking $\int_{v\in g_{-}^{-1}(\left\{ y\right\} )}\omega(v)g_n^{\prime}(v)dv$
out from the integral of $\int_{x\in g_{+}^{-1}(\left\{ y\right\} )}\,\cdot\;dx$
and $\int_{x\in g_{+}^{-1}(\left\{ y\right\} )}\frac{1}{\int_{v\in g_{+}^{-1}(\left\{ y\right\} )}dv}dx=\frac{\int_{x\in g_{+}^{-1}(\left\{ y\right\} )}dx}{\int_{v\in g_{+}^{-1}(\left\{ y\right\} )}dv}=1.$

Finally, by the definition of $\mathcal{Y}^{-}$ in \eqref{eq:Y_minus},
we have
\begin{equation}
\int_{\mathcal{X}^{-}}\omega(x)g_n^{\prime}(x)dx=\sum_{\mathcal{Y^{-}}}\int_{x\in g_{-}^{-1}(\left\{ y\right\} )}\omega(x)g_n^{\prime}(x)dx.\label{eq:w_2}
\end{equation}
Combining \eqref{eq:tilde_w_2} and \eqref{eq:w_2} yields the equivalence
result.

Second, we show the nonnegativity result, $\tilde{\omega}(x)\geq0\;\forall x\in\mathcal{X}$ for all. By definition \eqref{eq:induced_mu}, we have 
\begin{align*}
\int_{v\in g_{+}^{-1}(\left\{ g_n^{\prime}(x)\right\} )}\omega(v)dv & =\mu_{y^{+}}(\{g_n^{\prime}(x)\}),\\
\int_{v\in g_{-}^{-1}(\left\{ g_n^{\prime}(x)\right\} )}\omega(v)dv & =-\mu_{y^{-}}(\{g_n^{\prime}(x)\}).
\end{align*}
The conclusion follows by Condition\eqref{eq:measure dominance} and
the definition of $\tilde{\omega}_n(x)$ in \eqref{eq:step_w}.

\noindent \emph{Step 3.}
Now, we apply
Lebesque's Dominated Convergence Theorem (see, e.g., Theorem 1.34 of \citeauthor{rudin1987}, \citeyear{rudin1987}) to conclude the proof.
For Step 3, we only need to show:
\begin{itemize}
\item[(i)] There exist a function $m(\cdot)\in L^1(\mathcal{X})$ such that $|\tilde{\omega}_n(x)g_n^{\prime}(x)|\leq m(x)$ for  all $x\in \mathcal{X}$;
\item[(ii)] $\lim_{n\to\infty}\tilde{\omega}_n(x)g_n^{\prime}(x)=\tilde{\omega}(x)g^{\prime}(x)$.
\end{itemize}

For Condition (i), we can choose an appropriate sequence of discretized functions $g_n^{\prime}(\cdot)$, such that  $|g_n^{\prime}(x)|\leq 2|g^{\prime}(x)|$ holds for all $x\in \mathcal{X}$ and all $n\in \mathbb{N}$. By the continuity of $\omega(\cdot)$, the definition of $\tilde{\omega}_n(\cdot)$, and \eqref{eq:measure dominance} we have $\tilde{\omega}_n(\cdot)\leq \sup_{x\in\mathcal{X}}\omega(x)$. Condition (i) follows from choosing $m(x)=\sup_{x\in\mathcal{X}}\omega(x)\cdot 2|g^{\prime}(x)|$, the continuity of $g^{\prime}(\cdot)$, and the compactness of $\mathcal{X}$.

For Condition (ii), following the definitions of $ \tilde{\omega}_n(\cdot)$ and $\tilde{\omega}(\cdot)$, i$\lim_{n\to\infty} \tilde{\omega}_n(x)=\tilde{\omega}(x)$ holds for any $x\in\mathcal{X}$. Then,  Condition (ii) follows from $\lim_{n\to\infty} g_n^{\prime}(x)=g^{\prime}(x)$.

The conclusion of the proposition follows by applying Step 3 to Step 2.
\end{proof}

\bibliographystyle{apalike}
\bibliography{biball}
 	
\end{document}